\documentclass[%
 reprint,
 amsmath,amssymb,
 aps,nofootinbib,
  twocolumn,
]{revtex4-1}
\usepackage[usenames, dvipsnames]{color}
\usepackage[autostyle]{csquotes}
\usepackage{hyperref}
\usepackage{color}
\usepackage{graphicx}
\usepackage{dcolumn}
\usepackage{bm}
\usepackage{enumerate}
\usepackage{ulem}

\definecolor{MyDarkBlue}{rgb}{0,0.1,0.7}
\hypersetup{colorlinks,breaklinks=true,
  urlcolor={blue},citecolor={MyDarkBlue},linkcolor={magenta}}

\begin{document}

\preprint{APS/123-QED}

\title{Ricci-Determinant Gravity: Dynamical Aspects and Astrophysical Implications}
\author{Hemza Azri$^{1}$}
 \email{hmazri@uaeu.ac.ae; hemza.azri@cern.ch}
\author{K. Yavuz Ek\c{s}i$^{2}$}
\email{eksi@itu.edu.tr}
\author{Canan Karahan$^{2,3}$}
\email{ckarahan@itu.edu.tr}
\author{Salah Nasri$^{1,4}$}%
 \email{snasri@uaeu.ac.ae; salah.nasri@cern.ch}
\affiliation{$^{1}$Department of Physics, United Arab Emirates University, UAE
}%
\affiliation{$^{2}$Physics Engineering Department, Istanbul Technical University, 34469 Maslak, Istanbul, Turkey}%
\affiliation{$^{3}$National Defence University, Turkish Naval Academy, Department of Basic Sciences, Tuzla-Istanbul
}%
\affiliation{$^{4}$International Center for Theoretical Physics, Trieste, Italy}%

\date{\today}

\begin{abstract}

  The Palatini gravitational action is enlarged by an arbitrary function $f(\bm{X})$ of the determinants of the Ricci tensor and the metric, $\bm{X}=|\textbf{det}.R|/|\textbf{det}.g|$. The resulting Ricci-determinant theory exhibits novel deviations from general relativity. We study a particular realization where the extension is characterized by the square-root of the Ricci-determinant, $f(\bm{X})=\lambda_\text{Edd}\sqrt{\bm{X}}$, which corresponds to the famous Eddington action. We analyze the obtained equations for perfect fluid source and show that the affine connection can be solved in terms of the energy density and pressure of the fluid through an obtained disformal metric. As an application, we derive the hydrostatic equilibrium equations for relativistic stars and inspect the significant effects induced by the square-root of the Ricci tensor. We find that an upper bound on $\lambda_{\rm Edd}$, at which deviations from the predictions of general relativity on neutron stars become prominent, corresponds to the hierarchy between the Planck and the vacuum mass scales. The Ricci-determinant gravity that we propose here is expected to have interesting implications in other cosmological domains.

\end{abstract}

\maketitle

\section{\label{sec:intro}{Introductory remarks and motivation}}
\label{sec: introduction}
The new advances in cosmological observations have indicated that the standard model of cosmology, which stands on general relativity (GR), necessarily requires extensions \cite{supernova_ia}. The missing mass problem (or the dark matter), the luminosity of type Ia supernovae at large distances indicating an accelerated expansion of the Universe (the dark energy problem) and the lack of a consistent quantum theory of gravity are some of the facts that bespeak going beyond GR. This has led to numerous attempts to modify GR in various contexts including the familiar $f(R)$ gravity and others \cite{review_on_mog, ishak}. 

GR is purely a metric theory of gravity and these modifications are often formulated in this formalism. In the Palatini formalism the metric and the connection are assumed to be totally independent. Interestingly, the Palatini version of the Einstein-Hilbert action involves only the first derivatives of the connection and leads to the Einstein field equations without requiring an additional boundary term as in the standard action of GR which contains the second derivative of the metric \cite{palatini_action}.
In the last decades, extensions through the Palatini (or metric-affine) formulation have gained much interests \cite{extension_palatini}.  On the other hand, one of the most important consequences of this formulation is that it is free of ghosts that are usually present in metric theories of gravity due to the higher order equations for the metric.

Extensions of GR \textit{\`a la} Palatini, which are called generalized Palatini theories,  are usually constructed in terms of the powers of the traces of the Ricci tensor, i.e., generic functions of the form $f(g^{\mu\nu}R_{\mu\nu}, R^{\mu\nu}R_{\mu\nu})$ \cite{extension_palatini}. Despite being the standard way of building the gravitational theories, polynomial terms formed by contractions of the curvature (mainly the Ricci tensor) are not the only objects one can use in constructing invariant actions. In this respect, the determinant of the Ricci tensor also stands viable for the gravitational theories. Indeed, Eddington gravity as well as Eddington-inspired-Born-Infeld theories are modeled by the determinant of the Ricci tensor \cite{banados, lavinia-olmo}. Therefore, at least from a theoretical perspective, there is no rationale that prevents incorporating the determinant of the Ricci tensor into the Palatini formalism, given that it also is a Ricci-based object.

Motivated by these statements, in this paper, we enlarge the Palatini action by the determinant of the Ricci tensor. The extension will involve an arbitrary function $f(\bm{X})$ where the scalar $\bm{X}$ incorporates the determinant of the Ricci tensor. Since the latter is not an invariant scalar under general coordinate transformations, we also introduce the scalar density formed by the determinant of the metric tensor so that the new invariant quantity reads $\bm{X}=|\textbf{det}.R|/|\textbf{det}.g|$. This quantity has been used recently to enlarge Eddington gravity and incorporate matter \cite{tau}. Therefore, $f(\bm{X})$ is, indeed, a metric-affine scalar (\textit{\`a la} Palatini) but formed by scalar densities rather than the traces of the curvature tensor. The purpose of this article is to determine the role of the Ricci-determinant in the Palatini formulation of gravity, and reveal its dynamical effects that may not be present in theories with traces and powers of the Ricci tensor. Nonetheless, generalized Palatini theories can be improved by Ricci-determinant functions $f(\bm{X})$ unless a fundamental symmetry prohibits them.

 We derive the extended gravitational field equations of this \textit{Ricci-Determinant} theory by varying the total action with respect to the metric and the connection independently (see \S \ref{sec:ricci_determinant}). The simplest case of the theory occurs when $f(\bm{X})$ is merely constant. This only improves the Palatini (Einstein-Hilbert) action by a cosmological constant term. We show, in \S  \ref{sec:eddington_case}, that an interesting and simple model arises from the general theory when $f(\bm{X})=\lambda_\text{Edd}\sqrt{\bm{X}}$, with $\lambda_\text{Edd}$ being a dimensionless constant. It turns out that in this case the extension will be described by only the square-root of the Ricci-determinant which coincides with the familiar Eddington action. We will focus on this model and study it in details for the following reasons: First, the effects of the Eddington term and the role it plays in this enlarged theory is worth exploring in its own. More interestingly, the Ricci-determinant in this case will arise only in the dynamical equation obtained from varying the action with respect to the connection. Hence, the Ricci tensor will be easily written in terms of matter and the equation for the connection becomes linear and easy to solve.      
 
As a relevant application, we study the stellar structure equations of the $f(\bm{X})=\lambda_\text{Edd}\sqrt{\bm{X}}$ model (see \S~\ref{sec:Stellar_structure}). We then solve these equations  numerically, for 4 different equations of state, to obtain the mass-radius relations of neutron stars. These mass-radius relations, when confronted with the most recent observational measurements, allow us to constrain the sole free parameter of this model, $\lambda_\text{Edd}$.

We then discuss the mass scale associated to the obtained constraint on $\lambda_\text{Edd}$ and conclude in section \ref{sec:conclusion}. Some details on deriving the TOV equation are given in the appendix.

   
\section{Ricci-determinant gravity}
\label{sec:ricci_determinant}
\subsection{Action and gravitational field equations}
\label{sec:general_action}

In what follows the spacetime is assumed to be endowed with a Lorentzian metric $g$ and an independent symmetric connection $\Gamma$. One extends the Palatini action as
\begin{eqnarray}
\label{general_action}
S =&&\int d^{4}x \sqrt{|\textbf{det}.g|}
\,\Big\{
\frac{M^{2}_{\text{Pl}}}{2}\Big(g^{\mu\nu}R_{\mu\nu}(\Gamma)-2\Lambda\Big) +L^{\text{M}}[g]
\,\Big\}
\nonumber \\
&&+\int d^{4}x \sqrt{|\textbf{det}.g|}\, f(\bm{X})~,
\end{eqnarray}
where $\Lambda$ is a constant, \enquote{$\textbf{det.}$} refers to the determinant, and $L^{\text{M}}[g]$ is the Lagrangian density of matter fields. In this work, we will assume that $L^{\text{M}}[g]$ does not depend on the connection $\Gamma$. 

The scalar $f(\bm{X})$ is an arbitrary function of the scalar $\bm{X}$ which in turn involves the determinant of the Ricci tensor. Since the determinant is a scalar density, the general covariance implies that the $\bm{X}$ must be described by the ratio of two determinants, namely  
\begin{eqnarray}
\bm{X} \equiv \frac{|\textbf{det}.R|}{|\textbf{det}.g|}~.
\end{eqnarray}

In this work we will consider only the symmetric part of the Ricci tensor. Therefore, $R_{\mu\nu}$ simply refers to $R_{(\mu\nu)}$ throughout the paper. To that end, action (\ref{general_action}) has the following properties
\begin{itemize}
    \item The first line describes the Palatini version of GR with matter sources where the connection and the metric are independent fields.
    \item The last term enlarges the Palatini action with an arbitrary functions of the Ricci-determinant, not by
    generic functions $f(g^{\mu\nu}R_{\mu\nu}, R^{\mu\nu}R_{\mu\nu})$ of the Ricci-traces. Nevertheless, the latter are also allowed as in generalized Palatini theories and can be included in our setup.  
    \item The Ricci tensor in the overall action (\ref{general_action}) is linear in the derivatives of the connection, therefore the principle of variation will lead to the gravitational field equations without requiring any additional boundary term like the standard (purely metric) Einstein-Hilbert action.  
\end{itemize}

Variation with respect to the metric tensor leads to the generalized Einstein field equations
\begin{eqnarray}
\label{eq_1}
R_{\mu\nu}(\Gamma)=&&
 \Lambda g_{\mu\nu}
+\kappa \left( T_{\mu\nu}^{\text{M}} -\frac{1}{2}g_{\mu\nu}g^{\alpha\beta}T_{\alpha\beta}^{\text{M}} \right)  \nonumber \\
&&+ \kappa \left(2\bm{X}f^{\prime}(\bm{X})-f(\bm{X}) \right)g_{\mu\nu},
\end{eqnarray}
where $T_{\mu\nu}^{\text{M}}=L^{\text{M}}g_{\mu\nu}-2\delta L^{\text{M}}/\delta g^{\mu\nu}$ is the standard energy-momentum tensor of matter, $\kappa =1/M^{2}_{\text{Pl}}$ and $f^{\prime}(\bm{X})=df/d\bm{X}$.

Clearly, deviations from the standard Palatini gravitational equations manifest through the last term that involves $f(\bm{X})$ and its derivative. In the absence of matter sources, $T_{\mu\nu}^{\text{M}}=0$, the above equation reads
\begin{eqnarray}
R_{\mu\nu}(\Gamma)=
\Big( \Lambda 
+ \kappa \left(2\bm{X}f^{\prime}(\bm{X})-f(\bm{X})\right) \Big)g_{\mu\nu}.
\end{eqnarray}
This accepts a vacuum solution (when $f(\bm{X})$ is constant) where the Ricci tensor is proportional to the metric and an effective cosmological constant. We will illustrate this case with a specific model in the following section.

The second field equation, namely the dynamical equation, is obtained from variation with respect to the connection which leads to
\begin{eqnarray}
\label{eq_2}
\nabla_{\alpha}\left(\sqrt{|\textbf{det}.g|}\,g^{\mu\nu} +\frac{2\bm{X}f^{\prime}(\bm{X})}{M^{2}_{\text{Pl}}} 
\sqrt{|\textbf{det}.g|}(R^{-1})^{\mu\nu}
\right)=0 \nonumber \\
\end{eqnarray}

There are some important remarks on this equation which describes the evolution of the arbitrary connection $\Gamma$. First, unlike the standard Palatini theories, we notice the emergence of the inverse of the Ricci tensor, therefore, the Ricci tensor itself must not vanish in the first place to guarantee the existence of solutions to this equation. As we shall see in the next section, this will require a nonzero cosmological constant. Second, since the Ricci tensor (hence its determinant) involves the derivative of the connection, this equation is highly nonlinear and its solution is generally not trivial. In this respect, one can follow the same procedure used in generalized Palatini theories based on the Ricci scalar and Ricci-squared terms when solving the previous equation \cite{extension_palatini}. This stands on rewriting the Ricci tensor, which is the source of the nonlinearity, in terms of the metric and the stress-energy of matter with the aid of equation (\ref{eq_1}). However, generally speaking, this procedure may also not be simple due to the presence of the determinants in the last terms of (\ref{eq_1}). Nonetheless, in the next part of the paper, we will propose a model where this procedure can be applied directly and lead to an exact solution for the connection in terms of the metric, the energy density and pressure of a perfect fluid.  

\section{Model with the square-root of the Ricci-determinant}
\label{sec:eddington_case}
As we have explained so far, one way to solving the dynamical equation (\ref{eq_2}) is to make it linear in the connection. In other words, one writes the Ricci tensor in terms of matter with the aid of equation (\ref{eq_1}). This could have been trivial if the last term in (\ref{eq_1}) was proportional to the Ricci tensor not to its determinant. A much simpler and interesting case is when this equation is free of the effects of $f(\bm{X})$, i.e., the model in which the Ricci-determinant term manifests through the equation for the connection (\ref{eq_2}) solely. This arises simply when the last term of (\ref{eq_1}) vanishes 
\begin{eqnarray}
2\bm{X}f^{\prime}(\bm{X})-f(\bm{X}) =0,
\end{eqnarray}
which is characterized by the function
\begin{eqnarray}
\label{case_of_square_root}
f(\bm{X})=\lambda_{\text{Edd}}\sqrt{\bm{X}}
\end{eqnarray}
where $\lambda_{\text{Edd}}$ is a dimensionaless constant.

Returning to the theory (\ref{general_action}), one notices that this case corresponds to the Palatini action enlarged by the Eddington action
\begin{eqnarray}
\label{eddington_action}
\lambda_{\text{Edd}} \int d^{4}x\, \sqrt{|\textbf{det}.R|}.
\end{eqnarray}

Needless to say, this action could have been proposed as an extension of Palatini theory even if it does not involve the metric. Interestingly, it appears now as a particular model of the Ricci-determinant theory (\ref{general_action}). 

Therefore, the gravitational field equations (\ref{eq_1}) and the dynamical equation (\ref{eq_2}) read 
\begin{eqnarray}
\label{grav_eqs_from_metric}
&&R_{\mu\nu}(\Gamma)=
\Lambda g_{\mu\nu}+
\kappa \left(T_{\mu\nu}^{\text{M}} -\frac{1}{2}g_{\mu\nu}g^{\alpha\beta}T_{\alpha\beta}^{\text{M}}\right) \\
&&\label{dynamical_equation}
\nabla_{\alpha}\left(\sqrt{|\textbf{det}.g|}\,g^{\mu\nu} +\frac{\lambda_{\text{Edd}}}{M^{2}_{\text{Pl}}} \sqrt{\left|\textbf{det}.R \right|}
\left( R^{-1}\right)^{\mu\nu} \right)=0 \nonumber \\
\end{eqnarray}
respectively. 

As expected from this model, unlike the general case described by equation (\ref{eq_1}), the square-root of the Ricci-determinant (the Eddington’s term) does not contribute to the right-hand-side of the field equations (\ref{grav_eqs_from_metric}). This particular feature is not present in the standard generalized Palatini theories because the extensions are usually invariant terms formed by tensors contracted by the metric itself. However, it is known that one can build
an invariant action from the square-root of any rank-two tensor such as the Ricci tensor without invoking the metric. In other words, if one is able to remove the metric from the last term of (\ref{general_action}), the gravitational equations (\ref{eq_1}) would not involve the Ricci-determinant. It turns out that this is possible only for the particular model (\ref{case_of_square_root}).  

Again, the emergence of the inverse of the Ricci tensor which then requires that the latter must not vanish. However, from equation (\ref{grav_eqs_from_metric}), the curvature vanishes for the purely vacuum case in which $T_{\mu\nu}=0$ and $\Lambda=0$. Hence, this requires that a nonzero cosmological constant is essential in the theory. Therefore, the vacuum case will be described here by $T_{\mu\nu}=0$ and $\Lambda \neq 0$. It is worth noting that apart from the issue behind the theoretical estimation of its value, nonzero cosmological constant is strongly suggested by cosmological observations \cite{supernova_ia, weinberg_ccp, demir_cc, azri_cc}.

In this vacuum case, the previous gravitational equations simply read  
\begin{eqnarray}
&&R_{\mu\nu}(\Gamma)=
\Lambda g_{\mu\nu}, \\ 
&&\nabla_{\alpha}\left[
\left(1+\frac{\lambda_{\text{Edd}}\Lambda}{M_\text{Pl}^{2}}\right)
\sqrt{|\textbf{det}.g|}\,g^{\mu\nu}
\right]=0.
\label{effective_cc}
\end{eqnarray}
Since the factor that appears in (\ref{effective_cc}) is only a constant, one can easily show that this system of equations describes GR with a rescaled cosmological term $\Lambda/(1+\lambda_{\text{Edd}}\Lambda/M_\text{Pl}^{2})$.

\subsection{Solving for the connection}
\label{sec:solving_connection}
Returning to the case with matter, we notice that equation (\ref{dynamical_equation}) is not linear in the connection $\Gamma$ since it incorporates the Ricci curvature. Hence, a direct solution in terms of the metric might be complicated. However, the Ricci tensor is eventually written in terms of matter fields thanks to the gravitational field equations (\ref{grav_eqs_from_metric}). Therefore, equation (\ref{dynamical_equation}) is now linear since the connection comes out only through the covariant derivative. To solve this equation analytically, we follow the same procedure in generalized Palatini theories and introduce an \enquote{auxiliary} tensor $h_{\mu\nu}(x)$ given in its matrix form as
\begin{equation}
\label{tensor_h}
\hat{\bm{h}} = \left(\sqrt{\textbf{det}.\hat{\bm{\mathcal{P}}}} \right) \hat{\bm{\mathcal{P}}}^{-1}\, \hat{g}, 
\end{equation}
where the quantity $\hat{\bm{\mathcal{P}}}$ incorporates matter via the Ricci curvature as
\begin{equation}
\label{matrix_p}
\hat{\bm{\mathcal{P}}} = \hat{\bm{I}} + \frac{\lambda_{\text{Edd}}}{M^{2}_{\text{Pl}}} \frac{\sqrt{|\textbf{det}.R|}}{\sqrt{|\textbf{det}.g|}}\, \hat{\bm{R}}^{-1}.
\end{equation}

Here, the curvature and its inverse are calculated from the gravitational equations (\ref{grav_eqs_from_metric}) which will be finally written in terms of matter solely. This completes the solution of the dynamical equation (\ref{dynamical_equation}). As an application, we study below the case in which matter manifests as a perfect fluid which describes matter in various astrophysical and cosmological domains.

\subsection{The case of perfect fluids}
\label{sec:perfect_fluids}

Here, we take the stress-energy tensor in terms of energy density $\rho$ and pressure $P$ as 
\begin{equation}
T_{\mu\nu} = (\rho+P)u_{\mu}u_{\nu} +Pg_{\mu\nu}.
\end{equation}
Hence, one can show that from equation (\ref{grav_eqs_from_metric}) the inverse of the Ricci curvature in a matrix form reads 
\begin{eqnarray}
\left( R^{-1} \right)_{\mu}^{\,\,\nu} =
\frac{2M^{2}_{\text{Pl}}}{(\rho-P)} \delta_{\mu}^{\,\,\nu}
+\frac{4M^{2}_{\text{Pl}}\left(\rho + P \right)}{\left(\rho -P\right)\left(\rho +3P\right)}\,u_{\mu}u^{\nu}
\end{eqnarray}
where we have neglected the cosmological constant.

Finally, in terms of its components, the matrix $\hat{\bm{\mathcal{P}}}$ in (\ref{matrix_p}) takes the form
\begin{eqnarray}
\label{final_P}
\mathcal{P}_{\mu}^{\,\,\nu}=&&\left\{1+\frac{\lambda_{\text{Edd}}}{2M^{4}_{\text{Pl}}}\,\sqrt{\left(\rho-P\right)\left(\rho+3P\right)} \right\}\delta_{\mu}^{\nu} \\ \nonumber
&&+\frac{\lambda_{\text{Edd}}}{M^{4}_{\text{Pl}}}\,\left(\rho+P\right) \sqrt{\frac{\rho-P}{\rho+3P}}\,u_{\mu}u^{\nu}.
\end{eqnarray}
from which one calculates its determinant
\begin{eqnarray}
\textbf{det}.\hat{\bm{\mathcal{P}}}= (1+\bm{a})^{3}(1+\bm{a}-\bm{b}),
\end{eqnarray}
and its inverse
\begin{eqnarray}
(\mathcal{P}^{-1})_{\mu}^{\,\,\nu}=&&
-\bm{b}(1+\bm{a})^{-1}(1+\bm{a}-\bm{b})^{-1}\,u_{\mu}u^{\,\,\nu} \nonumber \\
&&+(1+\bm{a})^{-1}\,\delta_{\mu}^{\,\,\nu},
\end{eqnarray}
where the functions $\bm{a}$ and $\bm{b}$ are given in terms of the energy density, pressure, and the parameter $\lambda_{\text{Edd}}$as
\begin{eqnarray}
\label{a}
&&\bm{a}=
\frac{\lambda_{\text{Edd}}}{2M^{4}_{\text{Pl}}} \sqrt{\left(\rho-P\right)\left(\rho+3P\right)} \\
&&\bm{b}=\frac{\lambda_{\text{Edd}}}{M^{4}_{\text{Pl}}}\left(\rho+P\right) \sqrt{\frac{\rho-P}{\rho+3P}}.
\label{b}
\end{eqnarray}
With these quantities, we finally find the exact form of the tensor $h_{\mu\nu}$ from (\ref{tensor_h}) in terms of the physical metric $g_{\mu\nu}$ as
\begin{eqnarray}
\label{final_general_metric}
h_{\mu\nu}= \sqrt{(1+\bm{a})(1+\bm{a}-\bm{b})}\, \, g_{\mu\nu}
-\bm{b}\sqrt{\frac{1+\bm{a}}{1+\bm{a}-\bm{b}}}\,\,u_{\mu}u_{\nu}. \nonumber \\
\end{eqnarray}

In short, the spacetime connection in this model is reduced to the Levi-Civita of the tensor field (\ref{final_general_metric}) which involves both physical metric and matter (energy density and pressure). Therefore, the new effects arise in the matter sector by bringing out nonlinear terms in the energy density and pressure into the gravitational field equations. It is worth noticing that relation (\ref{final_general_metric}) has the form of the so-called disformal transformation. This type of transformations have been studied and applied to various models including the relativistic modified Newtonian dynamics \cite{mond}. However, relation (\ref{final_general_metric}) is not an imposed transformation of the metric but a result of the theory.    

Another remarkable point is that functions $\bm{a}$ and $\bm{b}$ that form the new tensor field $h_{\mu\nu}$ are proportional to the inverse of the fourth power of the Planck mass. Therefore, in the regimes where the energy densities (or pressure) are less than $M^{4}_{\text{Pl}}$, one is able to consider only first order terms in $\bm{a}$ and $\bm{b}$. Hence, in this case the previous expression reads  
\begin{eqnarray}
\label{final_metric}
h_{\mu\nu} \simeq \left(1+\bm{a}-\frac{\bm{b}}{2}\right) g_{\mu\nu}
-\bm{b}\,u_{\mu}u_{\nu}.
\end{eqnarray}
Notice here the GR limit, $h_{\mu\nu} \rightarrow g_{\mu\nu}$, as the zeroth order. This is clearly compatible with the case $f(\bm{X})=0$ in action 
(\ref{general_action}). In spacetime regions where the energy density reaches the Planck density, i.e.\ mostly near singularities, one must consider the general solution (\ref{final_general_metric}).

In the following section, we will consider the gravitational equations (\ref{grav_eqs_from_metric}) with the Levi-Civita connection of (\ref{final_metric}). We will apply the resulting equations to a static spherically symmetric spacetime, and then derive the stellar structure equations, namely, the Tolman-Oppenheimer-Volkov equations that correspond to this model.

\subsection{Stellar structure equations}
\label{sec:Stellar_structure}
The field equations (\ref{grav_eqs_from_metric}) will now be adapted to a static spherically symmetric spacetime with a line element given in terms of the physical metric $g_{\mu\nu}(x)$ as
\begin{equation}
ds^{2} = -{\rm e}^{2\nu(r)}\, dt^{2} + {\rm e}^{2\lambda(r)}\,dr^{2}
+r^{2}\,d\theta^{2} + r^{2}\sin^{2}\theta \, d\phi^{2}
\end{equation}
where $\nu(r)$ and $\lambda(r)$ are functions of the radial coordinate.

One writes the gravitational equations (\ref{grav_eqs_from_metric}) with mixed indices as
\begin{eqnarray}
\label{grav_egts_mixed_indices}
R_{\mu}^{\,\,\nu}(h)= \kappa\left(T_{\mu}^{\,\,\nu}-\frac{1}{2}\delta_{\mu}^{\,\,\nu} T \right) \equiv \mathcal{T}_{\mu}^{\,\,\nu}~.
\end{eqnarray}
Notice here that the curvature is now given in terms of $h_{\mu\nu}$, hence, it will certainly involve the metric $g_{\mu\nu}$, the energy density and pressure of the perfect fluid thanks to expression (\ref{final_metric}). The set of equations arising from (\ref{grav_egts_mixed_indices}) which describe the evolution of the gravitational potentials are derived with some details in the appendix. Here we summarize the stellar structure in two main equations as follows
\begin{eqnarray}
\label{psi_prime_equation}
\frac{d \Psi}{dr} =&&\frac{\kappa(\rho + P)r^{2}}{2(r- 2m)} - \frac{\kappa(\rho +P)r^{3}}{4(r - 2m)}\left(\bm{a}^{\prime}-\frac{\bm{b}^{\prime}}{2}  \right) \nonumber \\
&& - \frac{\kappa(\rho + 3P)r^{2}}{4(r -2m)}\bm{b} -\frac{\bm{b}^{\prime}}{2} + \frac{r}{2}\left(\bm{a}^{\prime \prime}-\frac{\bm{b}^{\prime \prime}}{2}  \right)
\end{eqnarray}
where we have introduced the mass $m(r)$ representing the total mass within the coordinate radius $r$, and the potential function $\Psi(r)$ such that
\begin{eqnarray}
\label{psi_and_m}
{\rm e}^{2\lambda(r)}=\left(1- \frac{2m(r)}{r} \right)^{-1}, \quad \Psi(r) = \nu +\lambda~,
\end{eqnarray}
and the prime signs refer to the derivatives with respect to the coordinate $r$.

The second equation of the stellar structure takes the form
\begin{eqnarray}
\label{m_prime_equation}
\frac{d m}{dr} =&& \frac{\kappa \rho r^{2}}{2} + \left(r - \frac{3m}{2} - \frac{\kappa \rho r^{3}}{4} \right)\left(\bm{a}^{\prime}-\frac{\bm{b}^{\prime}}{2}  \right) \nonumber \\
&&+\frac{r^{2}}{2}\left(1 -\frac{2m}{r} \right)\left(\bm{a}^{\prime \prime}-\frac{\bm{b}^{\prime \prime}}{2}  \right) \nonumber \\
&&-\frac{\kappa (\rho + 3P)r^{2}}{8}\bm{b}~.
\end{eqnarray}

From both equations one can easily extract the GR-limit which is obtained at the zeroth order, i.e., when the functions $\bm{a}$, $\bm{b}$ and their derivatives are ignored. Hence, the crucial difference from GR is that the additional terms incorporates not only the nonlinear terms in the energy density and pressure but the derivative of the latter too. Therefore, one expects that the modified TOV equation cannot be linear in the derivative of the pressure as in the case of GR.

To derive the TOV equation, one turns to the conservation equation $\nabla_{\mu}T^{\mu\nu}=0$ which reads
\begin{eqnarray}
\frac{d P}{d r}&&=-(\rho + P)\,\nu^{\prime} \nonumber \\
&&=-(\rho + P)\Big\{\Psi^{\prime} -\left(1 -\frac{2m}{r}\right)^{-1}\frac{m^{\prime}}{r} \Big\} \nonumber \\
&&\,\,\,\,\,-(\rho + P)\left(1 -\frac{2m}{r}\right)^{-1}\frac{m}{r^{2}}~.
\end{eqnarray}
The first part of this equation that includes $\Psi^{\prime}(r)$ and $m^{\prime}(r)$ can be obtained now by combining the previous expressions (\ref{psi_prime_equation}) and (\ref{m_prime_equation}), and finally one obtains a modified TOV equation
\begin{eqnarray}
\label{tov}
\frac{d P}{d r}=&&
-\frac{(\rho + P)}{r(r-2m)}
\left(m + \frac{\kappa Pr^{3}}{2} \right) \nonumber \\
&&+ \frac{(\rho + P)}{2(r-2m)}\left(m + \frac{\kappa Pr^{3}}{2} \right)\left(\bm{a}^{\prime}-\frac{\bm{b}^{\prime}}{2}  \right) \nonumber \\
&&+(\rho + P) \bm{a}^{\prime}+ \frac{\kappa(\rho + P)(\rho +3 P)r^{2}}{8(r-2m)}\bm{b}.
\end{eqnarray}
The first line in this equation shows the GR-limit. One notices that the new effects come through the nonlinear terms of the energy density, pressure and its derivative. Therefore, solution to this equation cannot obtained trivially even for the simplest equations of state relating pressure to energy density. Nevertheless, in the following section we will solve this equation perturbatively by considering the known GR solution at the zeroth order.

\subsubsection{Constraints via neutron stars}
\label{sec:Neutron_stars:the_numerical_solution}

\begin{figure*}[t!]
\includegraphics[width=0.49\textwidth]{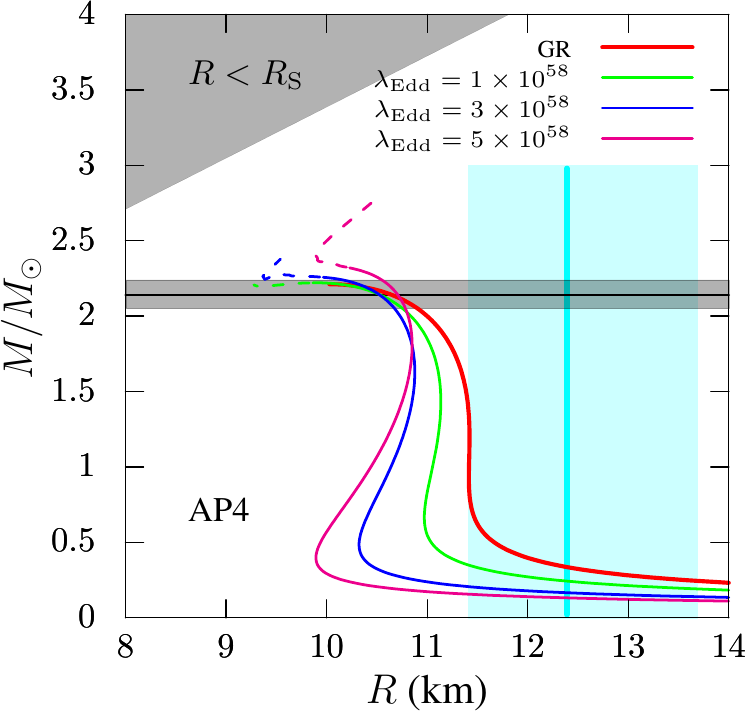}
\includegraphics[width=0.49\textwidth]{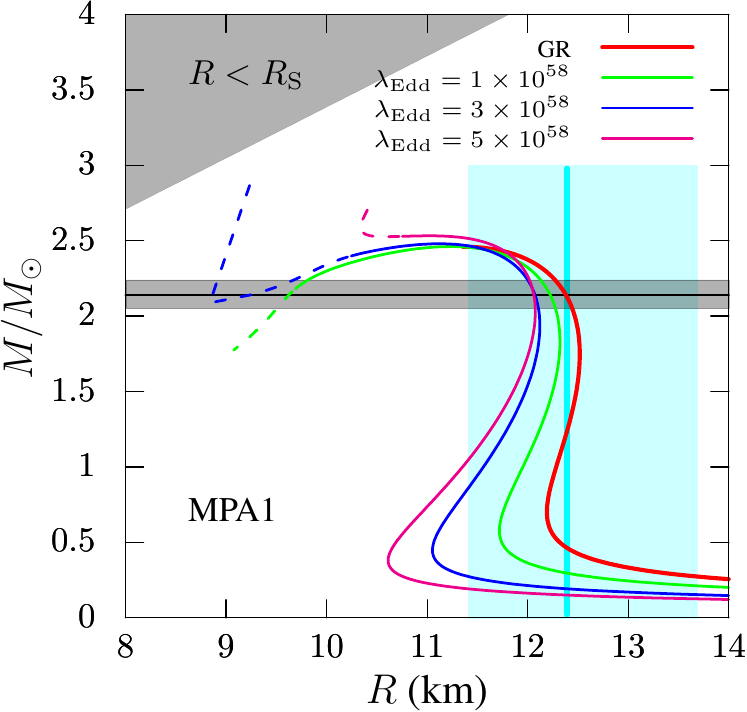} \\
\includegraphics[width=0.49\textwidth]{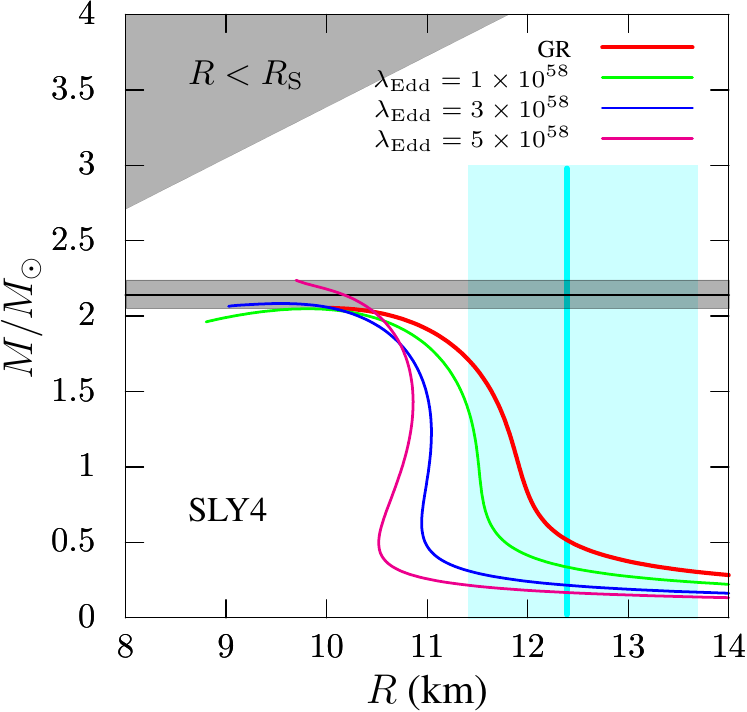}
\includegraphics[width=0.49\textwidth]{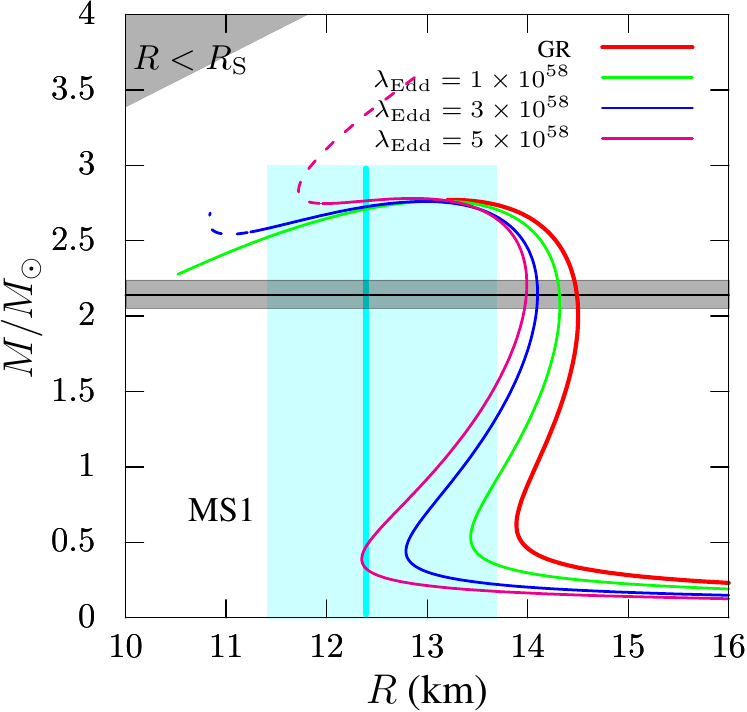}
\caption{Mass-radius relation of neutron stars in the Ricci-determinant model (\ref{case_of_square_root}). Each panel corresponds to a different equation of state.
The gray shaded region in each panel shows $R<R_{\rm S}\equiv 2GM/c^2$. The horizontal line shows the maximum measured mass $M = 2.14_{-0.09}^{+0.10}$ for a neutron star \cite{cro+20}. The cyan colored shaded region is the measured radius $R = 12.39_{-0.98}^{1.30}\,{\rm km}$ \cite{ril+21}.
}
\label{fig-MR}
\end{figure*}

In this section, we constrain the value of the parameter $\lambda_{\rm Edd}$ by demonstrating its effect on the mass-radius relation of neutron stars.

Neutron stars are very good laboratories for probing the strong gravity regime \cite{psaltis08}. The compactness and curvature of neutron stars at the surface \cite{dedeo03} and interior domain \cite{eksi+14} are orders of magnitude larger than the values probed in the solar system tests. Stellar mass black holes have slightly higher compactness and curvature at their horizon, but, since vacuum solutions in many theories of gravity are similar \cite{psa+08}, it is not possible to see the differences in the predictions of these theories from that of GR by the astrophysical observations of black holes.
Although the equation of state of neutron stars is not well constrained \cite{lat16,oze16}, the existing  mass and radius measurements \cite{cro+20,ril+21} can be used to constrain the free parameters of the gravity models, as an order of magnitude, for which deviations from GR becomes more prominent at higher curvatures. Accordingly, neutron stars have been used to constrain many models of gravity in the strong field regime 
\cite{ara+11,pani+11,del+12,yazad+14,cap+16,aka+18,ara+19}.

\subsubsection{Numerical method}

The hydrostatic equilibrium of a relativistic star is described by 
equations (\ref{m_prime_equation}) and (\ref{tov}).  To close the set of equations we need to supplement these equations with an EoS, $P=P(\rho)$. The EoS of dense matter prevailing at the cores of neutron stars is not strictly constrained by the nucleon scattering experiments. Several EoS with different assumptions about the nucleon-nucleon interactions and possible composition exist in the literature (see e.g.\ \cite{lat01}). We thus solve the equations for four different representative EoS to demonstrate the effect of the term (\ref{case_of_square_root}). As stated above, we use a perturbative method similar to the one employed in \cite{ara+11} where we calculate the higher order derivatives, such as $d^2 P/dr^2$, within GR.

Since equations (\ref{m_prime_equation}) and (\ref{tov}) are nonlinear and we use a complicated EoS, we need to obtain the solutions numerically.
To this end we employ the second order Runge-Kutta method (midpoint method) and use adaptive radial step-sizes which is adjusted according to the local mass and pressure gradients \cite{BPS}
\begin{equation}
\Delta r = 0.01 \left(\frac{1}{m}\frac{dm}{dr} - \frac{1}{P}\frac{dP}{dr} \right)^{-1}.
\end{equation}
This allows us to obtain sufficient radial resolution near the crust where the pressure gradient is large. 
We do not allow for the steps to grow larger than $10^3\,{\rm cm}$. 

We start from the origin ($r=0$) by choosing a central density $P_{\rm c}$, employ the boundary condition $m(0)=0$ and integrate outwards until we reach the surface i.e.\ where the pressure vanishes. This marks the radius of the star $R$ and the mass contained within ($m(R)$) is then the total mass of the star, $M$.

We then vary the central pressure within the range $3\times 10^{33} - 9\times 10^{36}\,{\rm dyne\, cm^{-2}}$ to obtain the corresponding mass and radius for each central pressure. We repeat this process for four different values of $\lambda_{\rm Edd}$ to obtain the mass-radius relation for each value of this parameter.

\begin{figure}[t!]
\includegraphics[width=\linewidth]{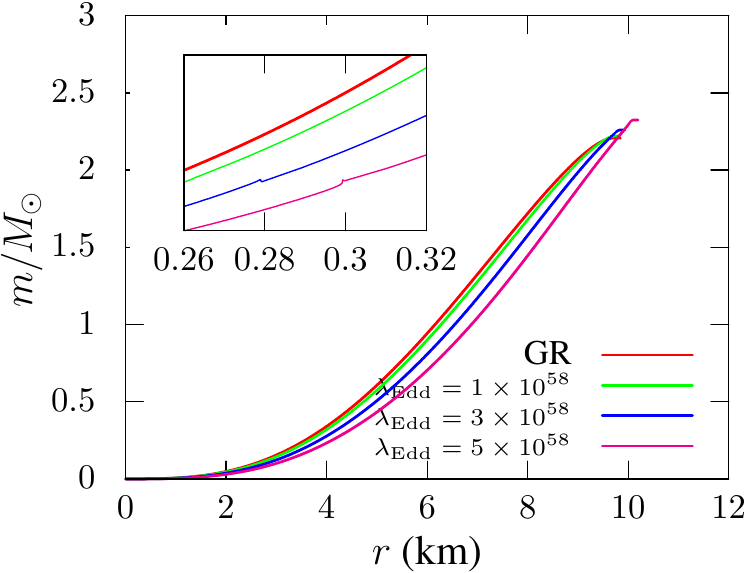}
\caption{The mass contained within radial coordinate $r$ for EoS AP4 with central pressure $P_{\rm c} = 2.05\times 10^{36}\,{\rm dyne\, cm^{-2}}$. All curves look normal, but the inset shows that there is a tiny region at which $dm/dr<0$ near $r=0.3\,{\rm km}$ i.e.\ close to the center. This situation, however, arises at very high densities corresponding to the near maximum of the M-R curves in \autoref{fig-MR}.}
\label{fig-mass}
\end{figure}
\subsubsection{Equations of state}

The above process is repeated for four different EoS: 
AP4 \cite{AP4}, 
SLY4 \cite{SLY},
MPA1 \cite{MPA1},
and MS1 \cite{MS1} which correspond to different assumptions about the composition and interactions of the dense nuclear matter. 
The order of magnitude of the constraint we obtain below will not change significantly if other EoS are used, but depends on the observational constraints on the mass and radius of neutron stars.

Instead of employing the tabulated EoS by interpolation, we used an analytical representation \cite{gun11}. This eliminates spurious oscillations in the radial structure solutions and mass-radius relations due to the presence of higher derivatives. 

\subsubsection{Results}

The mass-radius relations we obtained are depicted in \autoref{fig-MR} where each panel stands for a different EoS. We find that a choice of $\lambda_{\rm Edd} \sim 10^{58}$ leads to prominent changes in the mass-radius relation of neutron stars. 

The maximum mass measured from a neutron star is  $M = 2.14_{-0.09}^{+0.10}$ \cite{cro+20}. In order that an EoS and $\lambda_{\rm Edd}$ pair is eligible, the mass-radius curve should have a maximum value exceeding this. The radius of the same neutron star, PSR~J0740$+$6620, is measured to be $R = 12.39_{-0.98}^{1.30}\,{\rm km}$ \cite{ril+21}. This also can be used to constrain EoS and $\lambda_{\rm Edd}$ pairs, and here it clearly favours MPA1 with $\lambda_{\rm Edd} <1\times 10^{58}$. There are, of course, many other EoS that are compatible with these observations. Since our purpose is not to constrain the EoS of neutron stars, but to obtain an order of magnitude constraint on the value of $\lambda_{\rm Edd}$ we find it sufficient to present results only for four EoS. 

We find that the maximum mass increases with the parameter $\lambda_{\rm Edd}$. This allows one to obtain neutron stars with higher masses even with soft EoS. It is not possible to exploit this freedom to obtain arbitrarily large masses since $\lambda_{\rm Edd}$ is not entirely free, as we explain below.

Another constraint comes basically from the requirement that the mass within radial coordinate increases as one integrates outwards from the center ($dm/dr>0$). This is guaranteed in Newtonian gravity and general relativity while we find that this can not be taken for granted in the theory we consider here due to the presence of negative terms on the right hand side of Eqn.~(\ref{m_prime_equation}).
We find that at the highest densities there can be a narrow domain within the star where $dm/dr<0$ as shown in \autoref{fig-mass}. We plot the part of the M-R curves within which $dm/dr>0$ with dashed lines. This implies that the theory becomes incompatible with the existence of neutron stars at such high densities for  $\lambda_{\rm Edd} \sim 0.5\times 10^{58}$. 

It is worth noting here that the obtained constraint on $\lambda_{\rm Edd}$ is expected from the model at hand. Being large can be explained simply by the hierarchy between two relevant mass scales in the theory, namely, the Planck scale and a sub-eV scale of order the neutrino mass scale. While the former describes the gravitational mass scale, the latter characterizes the vacuum mass scale, i.e., the only and necessary source for Eddington gravity. In other words, one writes the gravitational action (\ref{eddington_action}) with $\lambda_{\rm Edd} = M^{2}_{\text{Pl}}/M^{2}_{0}$ and the above value implies $M_{0} \sim 0.1 \,\text{eV}$. This assures again that the vacuum energy (nonzero cosmological constant) is unavoidable feature in Eddington (square-root of the Ricci-determinant) theory of gravity \cite{induced_affine_inflation, affine_inflation, inducing_gravity_from_connections, separate_spaces, eddington_gravity_in_immersed_spacetime, azri_thesis, affine_dm, scalar_connection_gravity, entropy_production, azri_review, asymmetric}.

\section{Conclusion}
\label{sec:conclusion}

%
We have extended the Palatini action for gravity to involve any generic function of the scalar density $\bm{X}=|\textbf{det}.R|/|\textbf{det}.g|$. We have studied the new features and the dynamical aspects of this Ricci-determinant gravity. The obtained theory differs crucially from the familiar extensions of gravity that augment GR by quadratic or higher order (or even arbitrary) terms of the curvature such as $f(R)$ or other types of modified gravity theories.

We have focused on a particular model that arises from the general theory and coincides with the addition of the famous Eddington action that involves only the square-root of the Ricci-determinant. We have shown that, unlike the general case where solutions for the connection in terms of the metric are expected to be non-trivial, the equation for the connection in this model can be made linear and are easy to solve. To that end, following the same procedure in generalized Palatini theories, we have solved the equation for the affine connection in terms of the metric for a perfect fluid, and examined the novel contributions to the gravitational equations induced by the Eddington term which manifest as nonlinear functions in energy density and pressure of the fluid. 

These new contributions are found to be proportional to the square of the gravitational constant, therefore, we proceeded to a perturbative approach. By considering a static and spherically symmetric spacetime, we have been able to derive the associated hydrostatic equilibrium equations appropriate for astrophysical processes. We have then solved these equations for different EoS to obtain the radial structure and mass-radius relations. The observational constraints on the mass and radius of neutron stars, and the stability constraint $dm/dr>0$, allowed us to put an upper limit on the value of the free parameter of the model as $\lambda_{\rm Edd} \lesssim {\rm a~few}\times 10^{58}$. We find that this upper limit corresponds to the vacuum energy scale when the gravitational action is rewritten in terms of mass. 

The Ricci-determinant gravity, in its general framework, must be explored more. This is expected to reveal more interesting features when applied to other cosmological domains. Whether it can serve for an explanation of the missing mass without invoking nonluminous matter, or has a potential impact on the early universe singularity, are worth exploring and will be studied elsewhere.

\section*{acknowledgments}
The work of HA and SN is supported by the UAEU under UPAR Grant No. 12S004. The work of CK is supported by {\.I}T{\"U} BAP grant TAB-2020-42312.
\appendix

\section{Connection and curvature coefficients}
The components of the physical metric are given by $g_{\mu\nu}= \text{diag}(-{\rm e}^{-2\nu}, {\rm e}^{2\lambda}, r^{2}, r^{2}\sin^{2} \theta)$ which leads to the following components of the auxiliary metric (\ref{final_metric})
\begin{eqnarray}
&&h_{00} = \left(1 + \bm{a} +\frac{\bm{b}}{2}\right)g_{00} \\
&&h_{ij} = \left(1 + \bm{a} -\frac{\bm{b}}{2}\right)g_{ij}~.
\end{eqnarray}
The Levi-Civita connection of $h_{\mu\nu}$ is written as
\begin{eqnarray}
\label{gammah}
\Gamma_{\,\,\mu\nu}^{\lambda}(h)=
\frac{1}{2}h^{\lambda\alpha}
\left(\partial_{\mu}h_{\alpha\nu}+\partial_{\nu}h_{\mu\alpha}- \partial_{\alpha}h_{\mu\nu} \right)~.
\end{eqnarray}
Due to the smallness of the parameters $\bm{a}$ and $\bm{b}$ (see (\ref{a})-(\ref{b})), only linear terms will be considered in the following calculations. First, the connection (\ref{gammah}) has the nonzero coefficients
\begin{eqnarray}
&&\Gamma_{\,\,00}^{r} =\frac{1}{2}{\rm e}^{(\nu -\lambda)}
\left(2\nu^{\prime} +\bm{a}^{\prime} +\frac{\bm{b}^{\prime}}{2} + 2\nu^{\prime}\bm{b} \right) \\
&&\Gamma_{\,\,0r}^{0} =\frac{1}{2}
\left(2\nu^{\prime} +\bm{a}^{\prime} +\frac{\bm{b}^{\prime}}{2}\right) \\
&&\Gamma_{\,\,rr}^{r} =\frac{1}{2}
\left(2\lambda^{\prime} +\bm{a}^{\prime} -\frac{\bm{b}^{\prime}}{2}\right) \\
&&\Gamma_{\,\,\theta\theta}^{r} =-\frac{1}{2}{\rm e}^{-2\lambda}
\left(2r +\left(\bm{a}^{\prime} -\frac{\bm{b}^{\prime}}{2}\right)r^{2} \right)\\
&&\Gamma_{\,\,\phi\phi}^{r} =-r \sin^{2}\theta\, {\rm e}^{-2\lambda}
\left(1 +\frac{1}{2} \left(\bm{a}^{\prime} -\frac{\bm{b}^{\prime}}{2}\right)r \right) \\
&&\Gamma_{\,\,r\theta}^{\theta} =\frac{1}{r}
\left(1 +\frac{1}{2} \left(\bm{a}^{\prime} -\frac{\bm{b}^{\prime}}{2}\right)r\right) \\
&&\Gamma_{\,\,\phi\phi}^{\theta}=
-\sin\theta \cos \theta \\
&&\Gamma_{\,\,\theta\phi}^{\theta}= \cot \theta,
\end{eqnarray}
where the remaining components are obtained from the symmetric character of the connection, $\Gamma_{\,\,\nu\mu}^{\lambda}=\Gamma_{\,\,\mu\nu}^{\lambda}$.

Now, the Ricci tensor constructed from this connection
\begin{eqnarray}
R_{\mu\nu}=
\partial_{\lambda}\Gamma_{\,\mu\nu}^{\lambda}
- \partial_{\mu}\Gamma_{\,\lambda\nu}^{\lambda}
+\Gamma_{\,\rho\lambda}^{\rho}\Gamma_{\,\mu\nu}^{\lambda}
-\Gamma_{\,\mu\rho}^{\lambda}\Gamma_{\,\lambda\nu}^{\rho}
\end{eqnarray}
has the nonzero components
\begin{eqnarray}
\label{r00}
R_{00}=&&
{\rm e}^{(\nu-\lambda)}\left(\nu^{\prime \prime} +\nu^{\prime 2} -\nu^{\prime}\lambda^{\prime} +\frac{2\nu^{\prime}}{r} \right) \nonumber \\
&&+{\rm e}^{(\nu-\lambda)}\left(\nu^{\prime \prime} +\nu^{\prime 2} -\nu^{\prime}\lambda^{\prime} +\frac{2\nu^{\prime}}{2} \right)\bm{b} \nonumber \\
&&+\frac{1}{2}{\rm e}^{(\nu-\lambda)}
\left(3\nu^{\prime}- \lambda^{\prime} +\frac{2}{r} \right)\bm{a}^{\prime} \nonumber \\
&&+\frac{1}{4}{\rm e}^{(\nu-\lambda)}\left(\nu^{\prime}- \lambda^{\prime} +\frac{2}{r} \right)\bm{b}^{\prime} \nonumber \\
&&+\frac{1}{2}{\rm e}^{(\nu-\lambda)}\left(\bm{a}^{\prime \prime} +\frac{\bm{b}^{\prime \prime}}{2} \right)
\end{eqnarray}
and
\begin{eqnarray}
\label{r_rr}
R_{rr}=&&-\nu^{\prime \prime} -\nu^{\prime 2} +\nu^{\prime}\lambda^{\prime} +\frac{2\lambda^{\prime}}{r} \nonumber \\
&&+\frac{1}{2}\left(3\lambda^{\prime}- \nu^{\prime} -\frac{2}{r} \right)\bm{a}^{\prime} \nonumber \\
&&-\frac{1}{4}\left(3\nu^{\prime}+ \lambda^{\prime} -\frac{2}{r} \right)\bm{b}^{\prime} \nonumber \\
&&-\frac{3}{2}\bm{a}^{\prime \prime} +\frac{1}{4}\bm{b}^{\prime \prime},
\end{eqnarray}
and
\begin{eqnarray}
\label{r_theta_theta}
R_{\theta\theta}=&&
1 + {\rm e}^{-2\lambda}\left(r \lambda^{\prime} -r\nu^{\prime} -1 \right) \nonumber \\
&&+\frac{1}{2}r^{2}{\rm e}^{-2\lambda}\left( \lambda^{\prime} -\nu^{\prime} -\frac{4}{r} \right)\bm{a}^{\prime} \nonumber \\
&&-\frac{1}{4}r^{2}{\rm e}^{-2\lambda}\left( \lambda^{\prime} -\nu^{\prime} -\frac{2}{r} \right)\bm{b}^{\prime} \nonumber \\
&&-\frac{1}{2}r^{2}{\rm e}^{-2\lambda} \left(\bm{a}^{\prime \prime} -\frac{1}{2}\bm{b}^{\prime \prime} \right)~.
\end{eqnarray}
The final component is $R_{\phi \phi}=R_{\theta\theta}\sin^{2}\theta$. With these components, one is now able to derive the main set of equations (\ref{psi_prime_equation}), (\ref{m_prime_equation}) and (\ref{tov}) as we shall do below. 

\section{Stellar structure equations}
We start with the field equations (\ref{grav_egts_mixed_indices}), examine the quantities $(1-\bm{b})R_{0}^{\,0}-R_{r}^{\,r}$ and $R_{\theta}^{\,\theta}$, and equate them to $(1-\bm{b})\mathcal{T}_{0}^{\,0}-\mathcal{T}_{r}^{\,r}$ and $\mathcal{T}_{\theta}^{\,\theta}$, respectively.
Using the curvature components (\ref{r00}) and (\ref{r_rr}), one finds
\begin{eqnarray}
(1-\bm{b})R_{0}^{\,0}-R_{r}^{\,r}=&&
-\frac{2}{r}{\rm e}^{-2\lambda}(\nu^{\prime} + \lambda^{\prime}) \nonumber \\
&&-\frac{1}{2}{\rm e}^{-2\lambda}\left(2\bm{a}^{\prime} - \bm{b}^{\prime} \right)(\nu^{\prime} + \lambda^{\prime}) \nonumber \\
&&-\frac{1}{2}{\rm e}^{-2\lambda}\left(\frac{2\bm{b}^{\prime}}{r} -2\bm{a}^{\prime \prime} + \bm{b}^{\prime \prime} \right)~.
\end{eqnarray}
In terms of the parameters $\Psi(r)$ and $m(r)$ defined by (\ref{psi_and_m}), one easily writes the last equality as
\begin{eqnarray}
(1-\bm{b})R_{0}^{\,0}-R_{r}^{\,r} =&& -\frac{2}{r}\left(1-\frac{2m}{r}\right)\Psi^{\prime} \nonumber \\
&&-\frac{1}{2}\left(1-\frac{2m}{r}\right)\left(2\bm{a}^{\prime} - \bm{b}^{\prime} \right)\Psi^{\prime} \nonumber \\
&&-\frac{1}{r}\left(1-\frac{2m}{r}\right)\bm{b}^{\prime} \nonumber \\
&&+\frac{1}{2}\left(1-\frac{2m}{r}\right)\left(2\bm{a}^{\prime \prime} - \bm{b}^{\prime \prime} \right)~.
\end{eqnarray}
With this expression, one now sets up the equality $(1-\bm{b})R_{0}^{\,0}-R_{r}^{\,r}=(1-\bm{b})\mathcal{T}_{0}^{\,0}-\mathcal{T}_{r}^{\,r}$, dividing both sides by $1-2 m/r$ and get the main equation (\ref{psi_prime_equation}). 

Second, from the curvature component (\ref{r_theta_theta}), we simply have
\begin{eqnarray}
R_{\theta}^{\,\theta} =&&
\frac{1}{r} + \frac{1}{r}{\rm e}^{-2\lambda}\left(r (\lambda^{\prime} -\nu^{\prime}) -1 \right) \nonumber \\
&&+\frac{1}{2}{\rm e}^{-2\lambda}\left( \lambda^{\prime} -\nu^{\prime} -\frac{4}{r} \right)\bm{a}^{\prime} \nonumber \\
&&-\frac{1}{4}{\rm e}^{-2\lambda}\left( \lambda^{\prime} -\nu^{\prime} -\frac{2}{r} \right)\bm{b}^{\prime} \nonumber \\
&&-\frac{1}{2}{\rm e}^{-2\lambda} \left(\bm{a}^{\prime \prime} -\frac{1}{2}\bm{b}^{\prime \prime} \right)~.
\end{eqnarray}
In terms of $\Psi(r)$ and $m(r)$ in (\ref{psi_and_m}), one can easily check that
\begin{eqnarray}
{\rm e}^{-2\lambda}(\lambda^{\prime} -\nu^{\prime}) =\frac{2m^{\prime}}{r}-\frac{2m}{r^{2}}
-\left(1-\frac{2m}{r} \right)\Psi^{\prime}~.
\end{eqnarray}
By substituting this into the last expression, we find
\begin{eqnarray}
R_{\theta}^{\,\theta} =&&
\frac{2m}{r^{2}}- \left(\frac{1}{2}-\frac{m}{r} \right)\left(\frac{2}{r} +\bm{a}^{\prime} -\frac{\bm{b}^{\prime}}{2} \right)\Psi^{\prime} \nonumber \\
&&+\left(\frac{m^{\prime}}{r} - \frac{m}{r^{2}} \right)\left(\bm{a}^{\prime}-\frac{\bm{b}^{\prime}}{2}\right) \nonumber \\
&&-\left(\frac{1}{2r} -\frac{m}{r^{2}} \right)\left(\bm{b}^{\prime} -4\bm{a}^{\prime} \right) \nonumber \\
&&-\left(\frac{1}{2} -\frac{m}{r} \right)\left(\bm{a}^{\prime \prime} - \frac{\bm{b}^{\prime \prime}}{2} \right)~.
\end{eqnarray}
Notice here that the term
\begin{eqnarray}
\left(\frac{2}{r} +\bm{a}^{\prime} -\frac{\bm{b}^{\prime}}{2} \right)\Psi^{\prime}
\end{eqnarray}
can be extracted from equation (\ref{psi_prime_equation}). Hence, by performing this step the equality $R_{\theta}^{\,\theta}=\mathcal{T}_{\theta}^{\,\theta}$ finally leads to our equation (\ref{m_prime_equation}).

\end{document}